\documentclass[a4paper,12p]{article}

\usepackage{graphicx}  
\usepackage{color}
\usepackage{hyperref}
\usepackage{setspace}

\begin{document}
\title{Estimating single molecule conductance from spontaneous evolution of a molecular contact}

\author{M Gil, T Malinowski, M Iazykov and H R Klein\\
CINaM UMR CNRS 7325, Aix Marseille Univ, F-13288, Marseille cedex 9, France}
\date{}
\maketitle

\begin{abstract} 
We present an original method to estimate the conductivity of a single molecule anchored to nanometric-sized metallic electrodes, 
using a Mechanically Controlled Break Junction (MCBJ) operated at room temperature in liquid.
We record the conductance through the metal / molecules / metal nanocontact while keeping the metallic electrodes at a fixed distance.
Taking advantage of thermal diffusion and electromigration, 
 we let the contact naturally explore the more stable configurations around a chosen conductance value. 
 The conductance of a single molecule is estimated from a statistical analysis of raw conductance and conductance standard deviation data for 
 molecular contacts containing up to 14 molecules. The single molecule conductance values are interpreted as time-averaged conductance 
 of an ensemble of conformers at thermal equilibrium.
\end{abstract}


\section{Introduction}
\label{intro}
Molecular electronics (i.e. electronic components based on a single molecule \cite{Song2011} or a molecular assembly) as been the focus 
of extensive research for more than a decade. However, despite the wide range of problems addressed, a basic question remains : 
what can we learn from the conductance of a single molecule, and what are the main factors that affect the magnitude of measured conductances 
\cite{Salomon2003}? 

Of the techniques used to study electrical conduction through molecules 
\cite{Akkerman2008}, Scanning Tunneling Microscopy (STM \cite{Binnig1982})  and 
Break Junction techniques \cite{Muller1992} are the most appropriate ways to measure electrical conduction through a single 
molecule \cite{Chen2007,Schwarz2014}. These two techniques make use of nanometric- or atomic-sized electrodes \cite{Agrait2003} to 
connect a single molecule and measure the current flowing through it. Despite the simplicity and unicity of the principle,
the results obtained often exhibit a large spread (see e.g. figure 15 of ref. \cite{Akkerman2008}), since single molecule 
conductance  strongly depends on experimental conditions \cite{Ulrich2006}.

STM in break junction mode (STM-BJ) is probably the easiest and the most widely used technique when dealing with single molecule 
conductance \cite{Xu2003a}. In an STM-BJ set-up, metallic junctions are repeatedly formed by indenting a 
metallic surface with a tip, and broken by stretching the contact, leading to nanometric-sized metallic electrodes. 
Working in an organic solution, properly functionalized molecules
can spontaneously bond to the electrodes, making it possible to measure the conductance of these molecular bridges. In these experiments 
, however, the molecular junction is continuously streched during measurements. This stretching induces a strain which reduces
the lifetime of the junctions \cite{Evans2001, Tsutsui2008, Alwan2013}, influencing the conformation of the molecules within the contact, 
and thus  their conductance (see e.g. \cite{Gonzalez2008,Kim2017}). This can be viewed as a drawback and makes it difficult to analyze 
conductance data \cite{Gonzalez2006}.
Moreover, in these molecular junctions, current is flowing through different parallel paths : the molecule(s) and a tunneling channel. 
When the distance between electrodes varies, the contribution of each path to the total conductance evolves, making it difficult to analyze 
conductance data. This has necessitated the development of dedicated analysis procedures \cite{Quan2015,Wu2017}.

An alternative measurement technique has been proposed by W. Haiss and co-workers using an STM \cite{Haiss2004}. 
This method consists in trapping molecules between an STM tip and a surface kept at a constant distance. The molecules, upon thermal motion,
spontaneously connect (deconnect) the two electrodes. Abrupt jumps in the tunneling current are clearly observed, attributed to molecular 
connection events, and single molecule conductance values are extracted unambiguously. This technique provides further insights into the 
conduction mechanism at the single molecule scale \cite{Haiss2006}. It highlights the temperature dependence of 
molecular conductance, governed by the conformer distribution of the molecules.

The present work reports an original technique, using a Mechanically Controlled Break Junction (MCBJ), and derived from the
above technique. The key idea is to take advantage of thermal diffusion and electromigration at room temperature to let 
the contact self-organize at the atomic scale, so that it naturally explores the more stable configurations around an 
average chosen conductance value. Similar ``random'' approaches have succesfully been applied to extract 
information from disordered or stochastic systems \cite{Krachmalnicoff2010, Malinowski2016}.
At the nanometric scale, tracking of diffusion processes gives rise to stochastic signals often in the form of random telegraphic signals 
. Unbiased statistical analysis of these signals can provide access to valuable information \cite{Armstrong2010, Brunner2014} without requiring 
underlying assumptions.

In the case of a molecular contact containing several molecules bridging two metallic electrodes, current is likely to vary in response  
to connection / deconnection events. If the contact contains only one type of molecule, all the 
individual events may be of the same amplitude, irrespective of the number of molecules forming the contact. 
In this situation, the conductance of a single molecule should be extracted from an analysis of these events. 
It should be noted that this quantity represents the conductance of a single molecule anchored to its metallic electrodes, 
thus including the intrinsic molecular conductance and the conductance of the contacts \cite{VUILLAUME2008}.

In the following, we present our experimental set-up and measurement principle. We fix the distance between the two 
metallic electrodes of the MCBJ and record the temporal evolution of the current flowing through it. Analyzing measurements 
performed on widely studied molecules, alkanedithiols, we show the pertinence and promise of this 
measurement technique by estimating the conductance of a single molecule from the spontaneous (thermal) evolution of molecular contacts 
composed of up to 14 molecules.
	  
\section{Methods}
\label{methods}
\subsection{Experimental setup}
Our set-up consists of a home-made MCBJ (described in \cite{Alwan2013}), and the related acquisition and  control electronics 
and informatics. The MCBJ was first introduced by Muller \cite{Muller1992}, based on 
an earlier design by Moreland and Ekin \cite{Moreland1985}. The basic principle consists in stretching a metallic wire by 
bending an elastic substrate supporting the wire with a mechanical actuator until the junction breaks, giving two 
separate electrodes. The electrode separation can then be adusted by a feedback loop  using the current flowing through the junction.
MCBJ exhibits excellent mechanical stability, which results from the reduction of the mechanical loop connecting one electrode to the other.
For this study, our MCBJ is operated at room temperature in organic solutions.

\begin{figure*}
\begin{center}
\includegraphics[width=70mm]{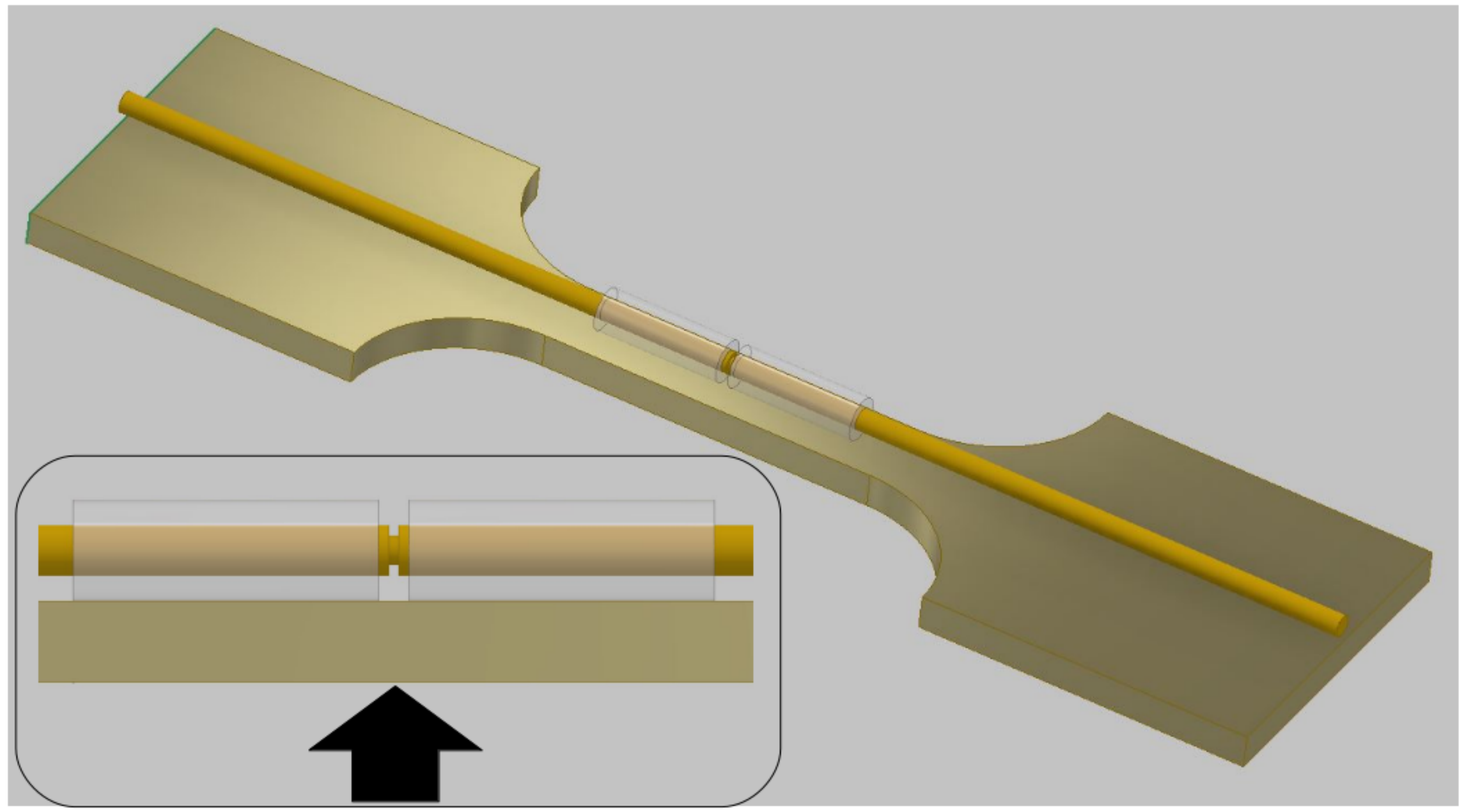}
\end{center}
\caption{ Schematic of the sample used for our MCBJ. The samples are made from Au wire glued into two quartz capillaries 
which are then glued onto a phosphorous bronze bending beam. As seen in the inset, the wire is notched in the empty space between 
the capillaries to initiate the breaking of the junction. See text for more details.
}
\label{fig : mcbj_sample}
\end{figure*}

Figure \ref{fig : mcbj_sample} represents a sample for our MCBJ. They are made from Au wire (250 $\mu$m diameter, 99.99\%, Goodfellow) glued 
into two quartz capillaries (fused silica, 1mm inner diameter, Vitrocom) which are then glued onto a 
phosphorous bronze bending beam. An optical glue (NOA61, Nordland) is used for both bondings. 
The wire is notched in the empty space between the capillaries to initiate the breaking of the junction.
The free-standing part of the Au wire is typically 200 $\mu$m for our samples, and the unfilled 
parts of the capillaries act as a reservoir for the organic solutions. Before use, samples are cleaned in a 
plasma cleaner (ATTO, Diener Electronic) operated with an air pressure of 0.4 mbar at a power of 30W for two minutes. 

The separation of the electrodes is controlled by a micrometer step motor (Z-825, Thorlabs) stacked-up with a 
piezoelectric actuator (sensitivity : 216 nm.V$^{-1}$). Motor and piezo are 
driven through an input/output board by a dedicated computer interface, which is also 
used for acquiring data and feedbacking. Because the wire is fixed, the actuator motion is demagnified, allowing 
accurate control of wire stretching. Taking into account a typical push:stretch ratio of 20:1 and the resolution of our 
16-bit DAC, one digit corresponds to less than $3\ pm$, which is ample for this work.
As the stability of the MCBJ is a critical parameter for this study, it is operated in the basement 
of the laboratory on an optical table to ensure optimal isolation from mechanical vibrations in a 
temperature-controlled environment (temperature variations below 1$^{\circ}$C on a 24 hour scale).
At low bias (typically, $V_{bias} \simeq 130\ mV$ for this study) and at 
room temperature, the drift of the electrodes is below 5 pm.s$^{-1}$ after one or two hours of operation.

The conductance is derived from the intensity that flows 
through the junction, measured using a home-made current/voltage converter with 
a logarithmic trans{-}conductance gain following the design proposed by U. D\"urig \cite{Durig1997}. 
This converter allows measurements from the 100pA range (noise level of 10pA on a 10 kHz bandwidth) up to the mA range, and is 
operated at a constant temperature of 20$^{\circ}$C.
Prior to measurements, the converter is carefully calibrated with a series of precision resistors 
(1,10,100 M$\Omega$ and 1 G$\Omega$  0.1\% resistors from Caddock). 
These calibration data are then used to calculate the current or conductance values corresponding to the output voltage. 
The junction, in series with a $1\ k\Omega$ ballast resistor to avoid saturation of the transimpedance amplifier,
is biased using a 6V lead battery. The voltage output of the transimpedance amplifier is recorded 
using a 16-bit ADC, operated at a sampling frequency of 32768 Hz. Fresh millimolar solutions of alkanedithiols 
(octanedithiols ODT $\geq$97\%, and pentanedithiol PDT 96\%, Aldrich) in mesitylene 
(98\%, Aldrich) are prepared and immediately used for measurements.

\subsection{Measurement and analysis protocol}

\begin{figure*}
\begin{center}
\includegraphics[width=140mm]{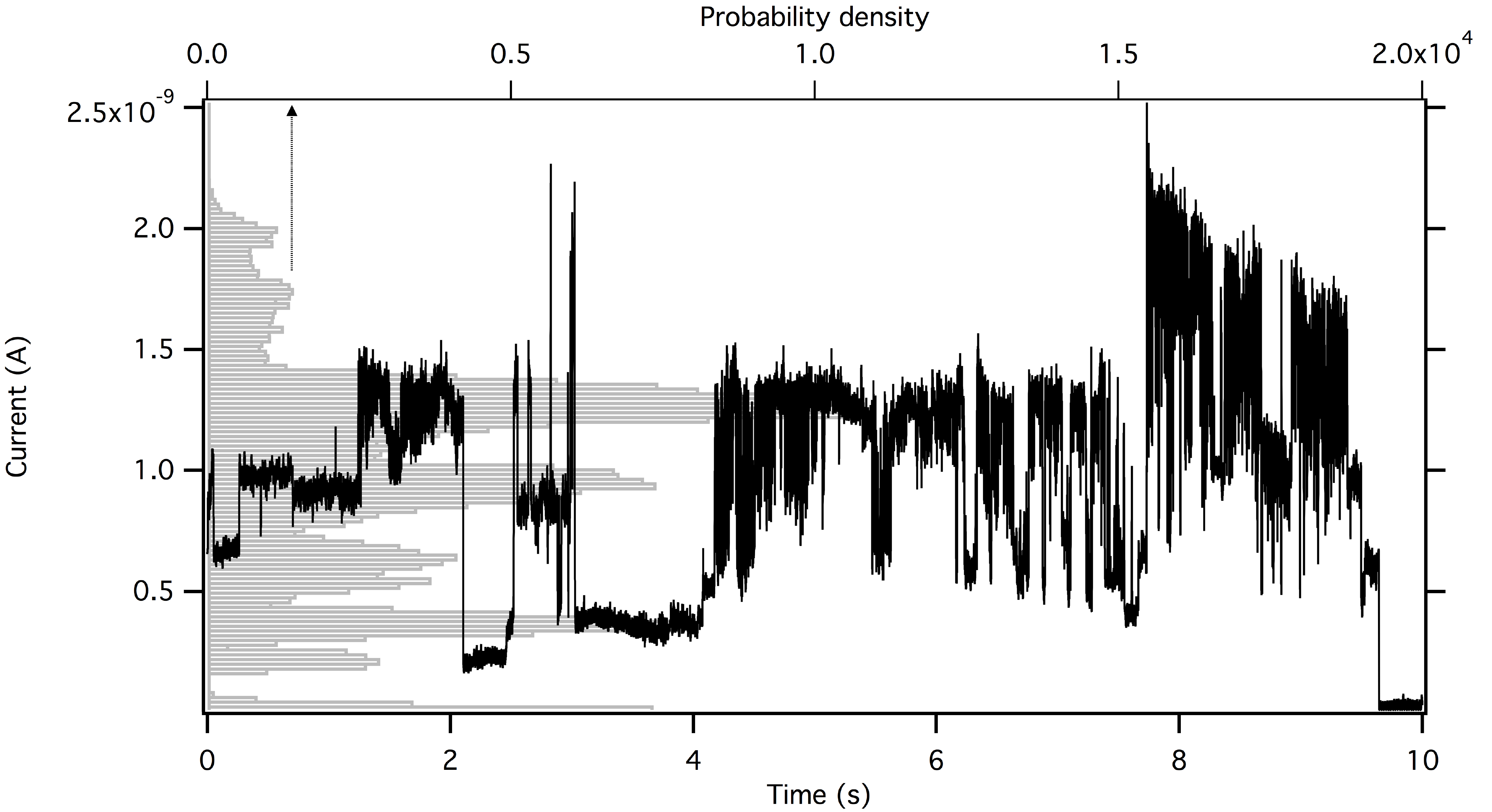}
\end{center}
\caption{ Spontaneous evolution of the current of a molecular contact formed from a millimolar octanedithiol (ODT) solution in mesitylene. 
Prior to recording, the junction is operated for 2 hours in pure solvent. The recording starts 5 minutes after the junction is refilled with 
the ODT solution. The electrode separation is fixed with a setpoint current of 0.5nA before the regulation feedback is disabled
(see text for more details). Current (black continuous line) is recorded, and a bias of 139mV is applied to the junction. 
A histogram of the current (gray) clearly indicates the most probable current values.
}
\label{fig : trace}
\end{figure*}

To ensure cleanliness of the contacts, samples are broken in pure solvent (mesitylene). 
They are operated for one or two hours at a typical setpoint of 200 pA before measurements begin, 
 to allow mechanical relaxation of the bending beam, and thus stabilization of the electrode distance. When certain that we observe no 
 noticeable current drift (beyond the noise level of the converter) on a minute range, we begin the measurements.

We feed the sample with 10 $\mu$L of solution and set a current setpoint in the sub nA range. This 
current imposes the distance between the gold electrodes, and after stabilization, the feedback loop controlling the electrode separation is disabled.

We then observe the temporal evolution of the current flowing through the contact. A connection or deconnection of a molecule between the 2 electrodes is 
expected to induce a sharp change (stepwise) in the current. If we do not observe such events, the feedback is enabled, and the setpoint 
is increased (reducing electrode separation). These operations are repeated until we observe abrupt current jumps. 

Taking advantage of thermal diffusion and electromigration, at room temperature, the nano-contact 
self-organizes at atomic level and naturally explores the more stable configurations around the average chosen conductance 
value. During these periods we record the temporal evolution of the current flowing through the contact. We operate the piezo actuator 
only if the current falls below  20 pA or rise above 10nA. The recording of the current is interrupted when the piezo actuator is active.

Figure \ref{fig : trace} illustrates a 10s recording of the current flowing through a junction filled with an octanedithiol solution and operated 
at a bias voltage of 139 mV. A random telegraphic signal is clearly observed, corresponding to numerous molecule connection / deconnection events
in the contact. A histogram of the current also clearly illustrates the discrete nature of current jumps and the 
most likely values of the current. This histogram and the others presented below are all constructed using the rule suggested by 
D.P. Doane \cite{Doane1976} for data binning, in order to be able to compare histograms constructed 
from datasets of different sizes\footnote{Current recordings typically contain from $10^7$ to $10^8$ samples.}. 

From the temporal evolution of the current (or conductance in the following) we can construct the distribution of the time spent in a given state by calculating the time lags 
between consecutive current jumps. Jumps are located by tracking variations in current standard deviation with a rolling window of 10 data points 
(equivalent to a cutoff of 0.3 ms for a sampling rate of 32768 Hz). We define a jump as a variation of more than 5 times the median 
standard deviation value of the current recording.

\section{Results}
\label{results}

Our main objective being to present the measurement technique, we conducted this work on widely studied saturated model molecules, alkanedithiols. 
We chose two alkanedithiol chains of different lengths : octane- and pentane- dithiols containing respectively 8 and 5 carbon 
atoms in their saturated backbones. These molecules bind to metallic electrodes via their terminal thiol groups.
Previous studies have shown unambiguously that transport through alkanedithiols can be approximated by a non-resonant 
tunelling process through a barrier whose width is determined by the distance between the terminal groups of the molecule \cite{Haiss2004}. 
We thus expect a higher conductance for the shorter pentanedithiol molecule than for the longer octanedithiol molecule, considering that both molecules have similar contact conductances.

\begin{figure*}
\begin{center}
\includegraphics[width=140mm]{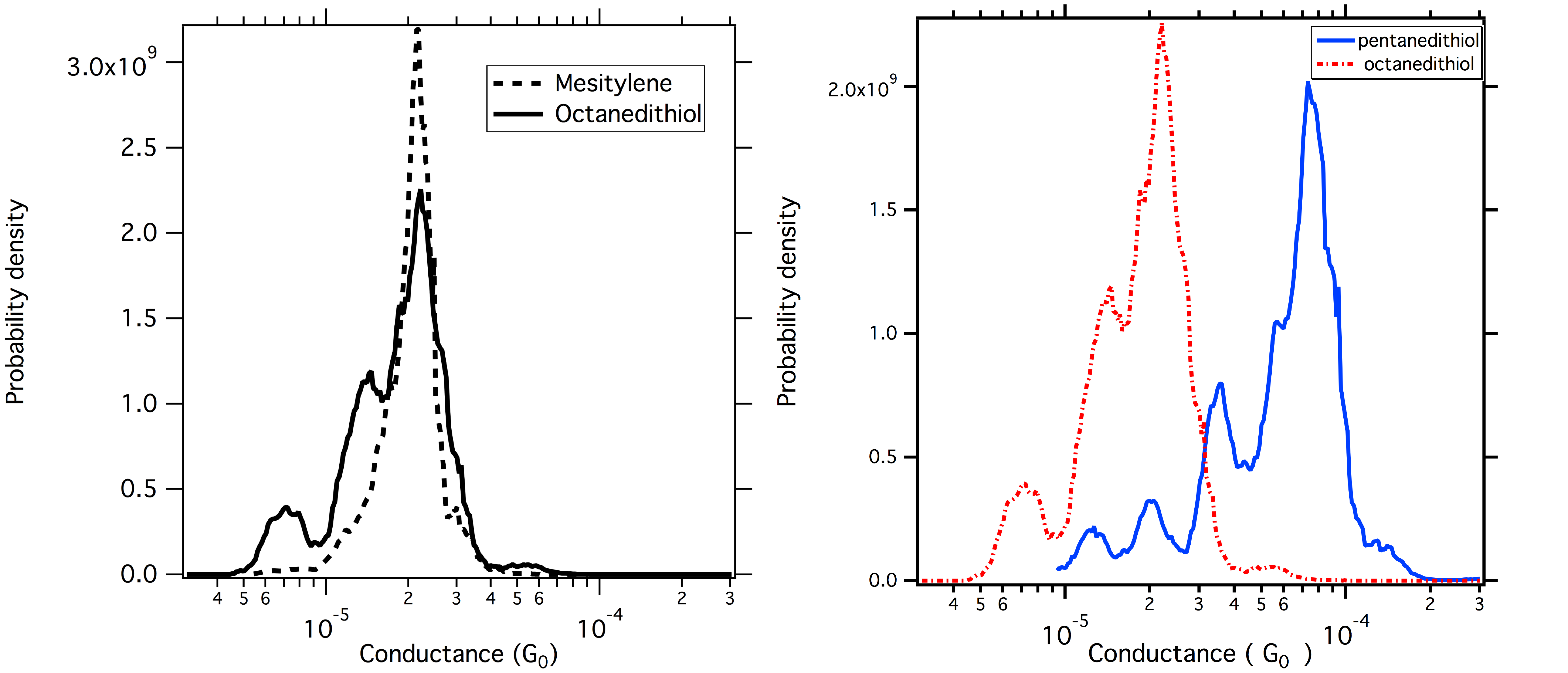}
\end{center}
\caption{
\textbf{Left :} Comparison of conductance histograms for a junction operated in pure mesitylene (black dashed line) and in 
 a millimolar solution of ODT in mesitylene (black continuous line). Bias voltage : 139 mV, setpoint current used to fix the electrode separation : 
 0.2nA. While the solvent's histogram does not exhibit particular features apart from random fluctuations around the setpoint current, it can be seen 
 that a junction containing ODT molecules explores more conductance values and exhibits peaks representing the most probable conductance values.
 \textbf{Right :}
Comparison of conductance histograms for a junction immersed in octane (online version red) and pentanedithiols (online version blue). 
Bias voltage : 139 mV, setpoint current 0.2nA for octanedithiols and 0.6nA for pentanedithiols. Each histogram exhibits similar features : 
series of evenly spaced peaks in conductance, which we attribute to the discrete variations in conductance 
related to the number of molecules in the contact.}
\label{fig : conductance_histogram}
\end{figure*}

In figure \ref{fig : conductance_histogram} (right), we compare conductance recordings for PDT and ODT millimolar solutions in mesitylene, with a bias voltage of 139mV, 
and setpoint currents of  respectively 0.2nA and 0.6nA for ODT and PDT. As explained above, these setpoints were chosen as the smallest 
setpoints where we observe current jumps. A higher setpoint means that electrode separation is smaller for PDT, which is in
agreement with the fact that  PDT molecules are shorter than ODT molecules.
 Both histograms exhibit similar features : a series of evenly spaced  peaks in conductance. 
 The measurements realized in pure mesitylene do not exhibit such features, with the exception of current fluctuations around the selected current setpoint
  (0.2 nA for a bias voltage of 0,139V, $G=1,9.10^{-5} G_0$), as seen on figure \ref{fig : conductance_histogram}(left).

\begin{figure*}	
\begin{center}
\includegraphics[width=140mm]{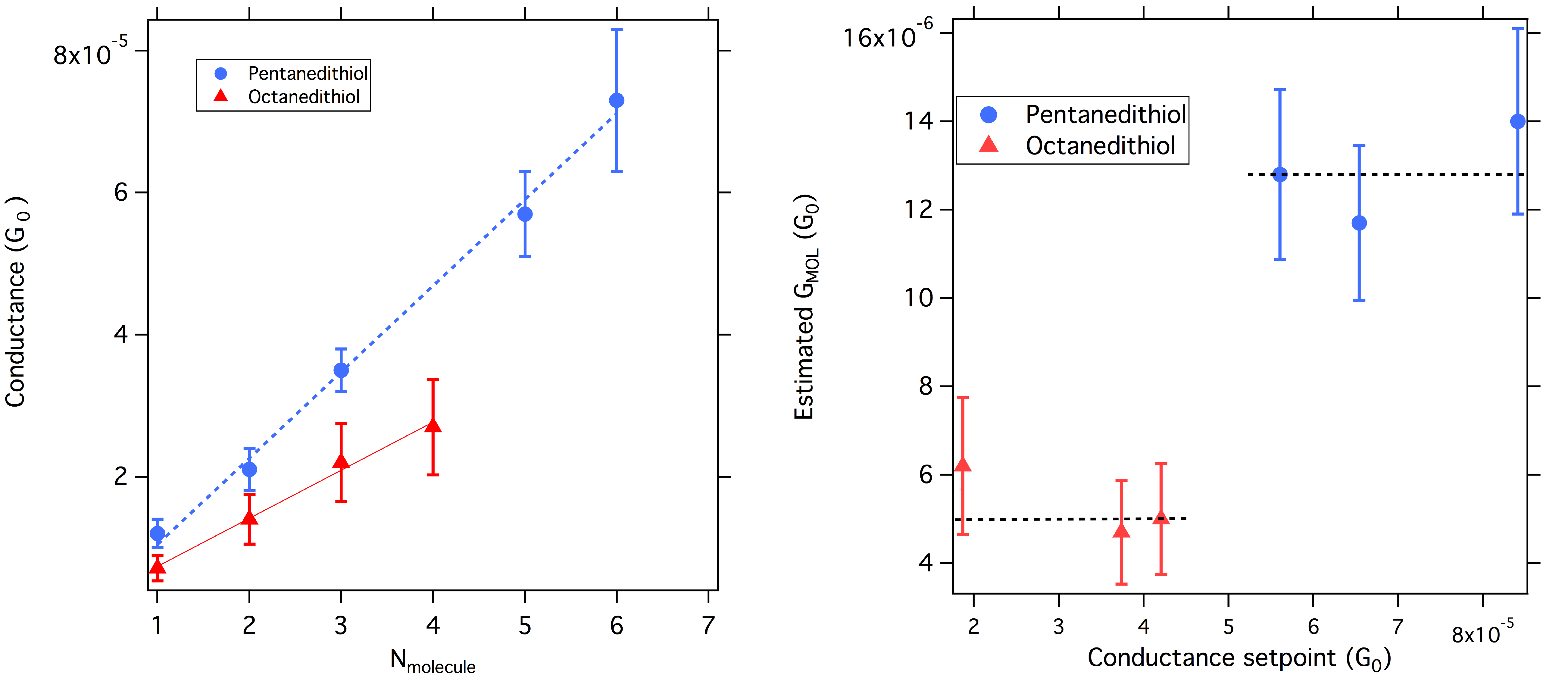}
\end{center}
\caption{\textbf{Left :} Conductance values of the peaks for the histograms of figure \ref{fig : conductance_histogram} versus a number of molecules. The lower 
values are attributed to a single molecule as we do not observe current jumps of lower amplitude. A linear fit of these curves gives a conductance
per molecule of $G_{PDT}= 1.22.10^{-5}\pm 5.10^{-7} G_0$ for PDT, and $G_{ODT}= 6.7.10^{-6}\pm 4.10^{-7} G_0$ for ODT.
\textbf{Right :} Estimation of conductance values for a single molecule for different conductance setpoints. Bias voltage : 139 mV. The 
estimated molecular conductance does not depend on the chosen setpoint current.
}
\label{fig : conductance_values}
\end{figure*}

Figure \ref{fig : conductance_values} (left) shows the conductance values corresponding to the clear peaks of both histograms 
of figure \ref{fig : conductance_histogram} on a graph where the abscissa is a number of molecules. The lowest conductance value is arbitrarily 
attributed to one molecule. The graph clearly shows that conductance values are fitted by a linear function $f(N_{MOL}) = G_{MOL} . N_{MOL}$, 
$ G_{MOL}$ being the conductance of a single molecule, and $N_{MOL}$ the number of molecules. 
From linear fits of these data points we extract two $G_{MOL}$ values : 
$G_{PDT}= 1.22.10^{-5}\pm 5.10^{-7} G_0$ for PDT, and $G_{ODT}= 6.7.10^{-6}\pm 4.10^{-7} G_0$ for ODT. 

Experiments were repeated for different setpoints for PDT and ODT. We do not expect strong dependence of the single molecule 
conductance on the chosen setpoint; the setpoint only fixes the initial configuration of the contact, and thus the number 
of molecules intially present. This is shown on figure \ref{fig : conductance_values}(right), where we plot 3 values of $G_{MOL}$ 
obtained from experiments following the same protocol at different setpoints for both PDT and ODT. These values are identical.

\begin{figure*}	
\begin{center}
\includegraphics[width=140mm]{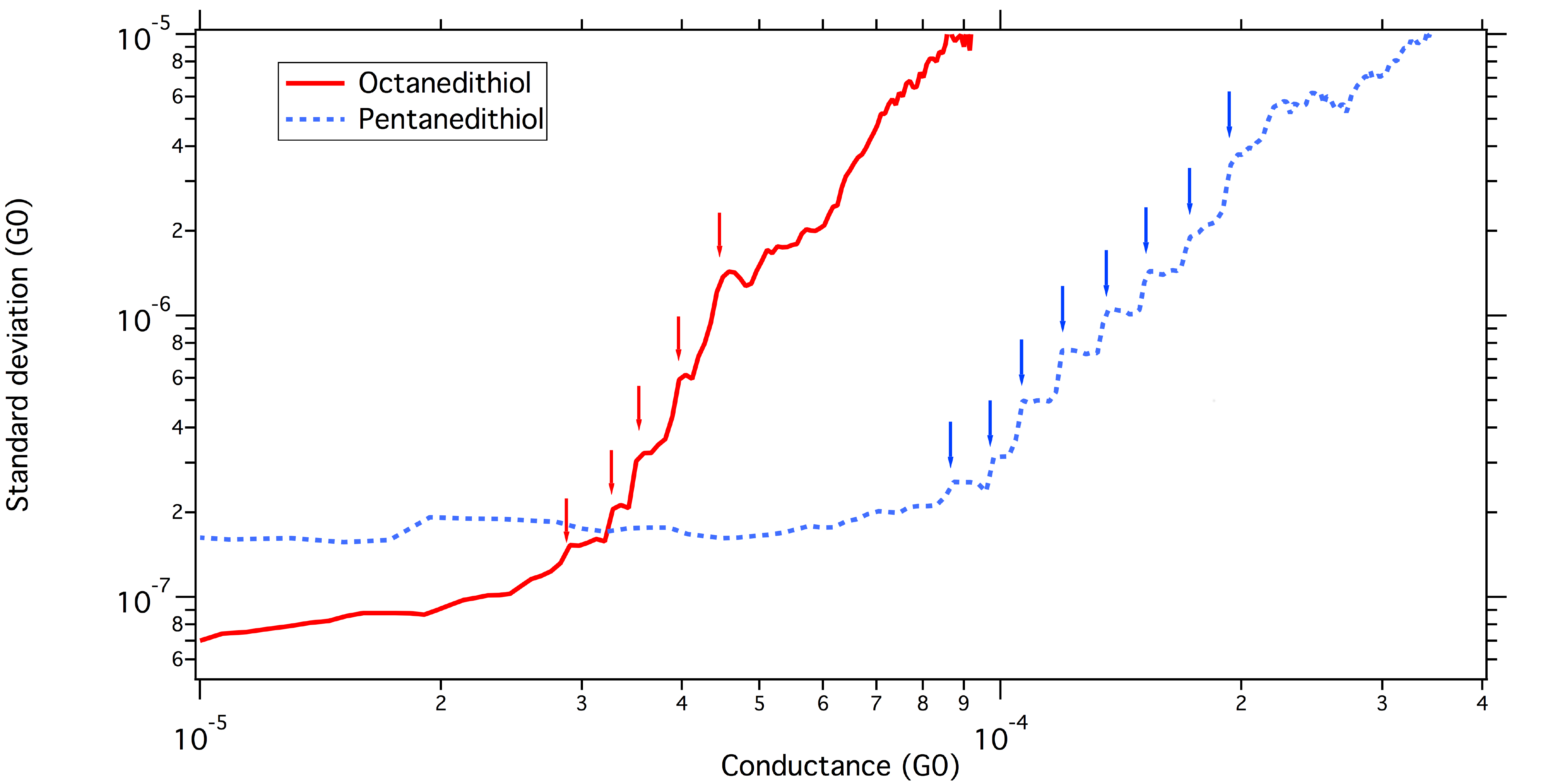}
\end{center}
\caption{Conductance standard deviation calculated for ODT (continuous line) and PDT (dashed line). Standard deviation data have been binned along the conductance
axis (see text for details), and median value of the standard deviation has been calculated for each bin. For both molecules, jumps 
in the standard deviation appear clearly. The spacing beween the jumps along the conductance axis is equal to individual conductance.}
\label{fig : conductance_std}
\end{figure*}

The discrete variations in conductance allow us to estimate single molecule conductance from a statistical analysis of raw 
conductance recordings. However, this evolution is rapidly smeared out by conductance fluctuations. It is therefore worth investigating whether 
current fluctuations can also provide information concerning the discrete evolution of the contact. If we make the assumption that the conductance 
fluctuation amplitude is related to the number of molecules acting as noise sources in the contacts, a discrete variation is expected.

To further explore these variations in conductance we estimate these fluctuations by calculating the standard deviation of the 
conductance of the contacts, integrated over the bandwidth of our recording 
set-up. Although it is not designed for accurate noise measurements, it still gives us interesting information.
Directly plotting the conductance standard deviation vs the conductance gives an informationless cloud of points. 
We applied to these data a binning procedure successfully used for another study \cite{Malinowski2016}. 
We divide standard deviation data in classes over the conductance axis, using Doane's rule. We then calculate the median 
standard deviation value for each class. The result of this analysis for the PDT and ODT experiments of figure \ref{fig : conductance_histogram} is 
shown in figure \ref{fig : conductance_std}. We can see a constant increase in standard deviation as the conductance increases. 
Despite the rough estimation method used, beyond the noise floor of our setup ( standard deviation larger than 10$^{-7}$G$_0$ ), 
jumps in standard deviation occur for a limited range of conductance : $3-5.10^{-5}G_0$ for 
ODT and $0.9-3.10^{-4}G_0$ for PDT. It is noteworthy that these jumps occur at evenly spaced conductance increases. 
Moreover, these values are consistent with those extracted from conductance histograms of figure \ref{fig : conductance_histogram}, but higher. 
Interestingly, when conductance jumps are smeared out by fluctuations, the dicrete nature of the molecular 
contacts is still visible in the fluctuations, although in a limited range. It allows us to perform a statistical analysis of the conductance 
of the contact on a larger set of observations; this is dicussed in the next section.

\section{Discussion}
\label{discussion}
The key idea behind this work is to estimate the conductance of individual molecules anchored to metallic electrodes 
when no external mechanical 
strain is applied. This is achieved by observing, in an MCBJ, the spontaneous evolution of a molecular contact formed 
between two metallic electrodes kept at a fixed distance, through the evolution of conductance.

In this situation, the reconfiguration of the molecular contact is thermally driven (see e.g.\cite{Brunner2014}). Many processes can 
alter the conductance of the contact : connection or deconnection of molecules, atomic rearrangements of the metallic 
electrodes and structural changes in the molecules. Rearrangements of electrodes \cite{Lynn1973}, and molecular conformational 
changes \cite{Vollhardt1987}, are expected to occur at a time scale below the nanosecond, which means that only averaged values will 
be measured within the limited bandwidth of our set-up. 
However, molecular connection in the contact is expected to occur at a millisecond time scale, given the chemical affinity of the 
thiol group to the gold electrodes \cite{Pensa2012}.

\begin{figure}	
\begin{center}
\includegraphics[width=60mm]{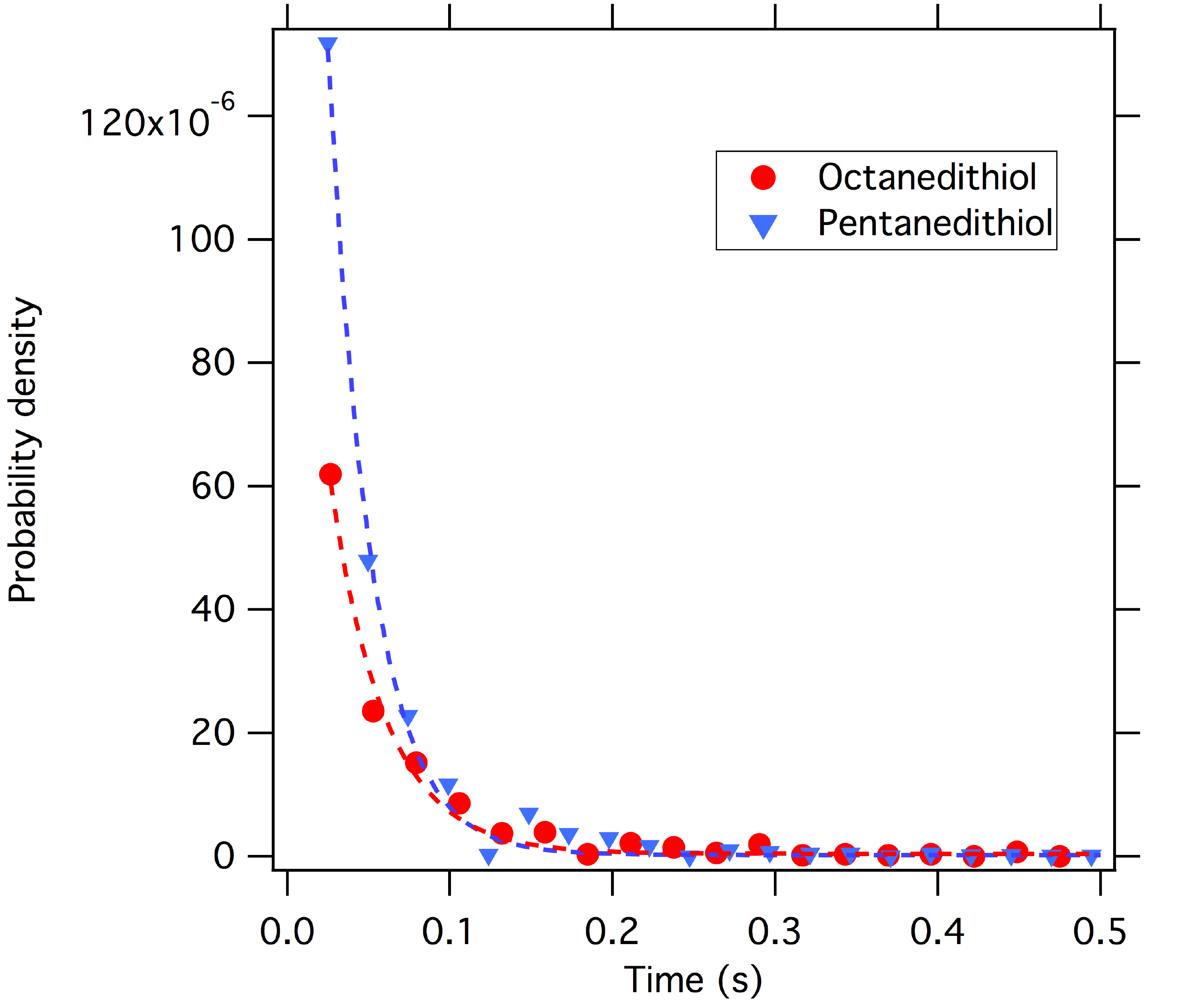}
\end{center}
\caption{Histogram of the time lag separating conductance jumps for pentane and octanedithiol in mesitylene. 
The mean time can be interpreted as the lifetime of one molecule in the contact. A fit (lines on plot) with an exponential function $A.\exp(-t/\tau)$ 
gives equivalent times $\tau_{PDT}=2.6.10^{-2} \pm 5.10^{-3}$ s for PDT, and $\tau_{ODT}=3.4.10^{-2} \pm 1.3.10^{-2}$ s for ODT.
}
\label{fig : time_lag}
\end{figure}

From the temporal analysis of current jumps shown in figure \ref{fig : time_lag}, 
we estimate mean time constants for connection / deconnection of molecules to be 30ms for both PDT and ODT. The mean lifetimes of 
connection events for the two studied molecules are very similar, since they are very similar and share the same anchoring group. 

We can thus state that these events are responsible for the abrupt current jumps that we systematically observe in molecular contacts. 
The current flowing through the electrodes comes from two channels : a tunneling channel whose conductance depends mainly on the 
electrode separation, and a molecular channel whose conductance depends on the nature, the number and possibly the conformation of the molecules.
With a constant distance between electrodes, the contribution of the tunneling channel does not vary, or only slowly due to thermal drift of the 
set-up.

By building a conductance histogram from current recording during the evolution of a contact, we obtain peaks indicating the most probable values. 
In our situation, these values should correspond 
to integer numbers of molecules within the contact, and are therefore expected to share a smallest common multiple, which we assign to the 
conductance of a single molecule. One of the main benefits of such an analysis lies in eliminating the constant (or slowly varying) 
tunneling contribution to net conductance, provided that several conductance values corresponding to different contact configurations 
are measured.

Thanks to the outstanding mechanical stability of our setup, we estimate the conductance of a single molecule to be 
$G_{PDT}= 1.28.10^{-5}\pm 0.21.10^{-5} G_0$ for PDT, and $G_{ODT}= 5.3.10^{-6}\pm 1.1.10^{-6} G_0$ for ODT. 
These values are in quantitative agreement with values published  for PDT and ODT, 
using a similar approach \cite{Haiss2004}. The fact that at least 5 evenly spaced conductance peaks are observed in 
the conductance histograms is a clear indication that the position of the electrodes remains fixed throughout 
the recording (typically 10 seconds to one minute). It is noteworthy that we found values  within the confidence intervals for 
both molecules (data not shown) at different biases (260 and 420 mV) . Within this bias range, there is likely to be no resonance with molecular levels 
of the alkanedithiols, and therefore a constant conductance value is expected.

\begin{figure}	
\begin{center}
\includegraphics[width=70mm]{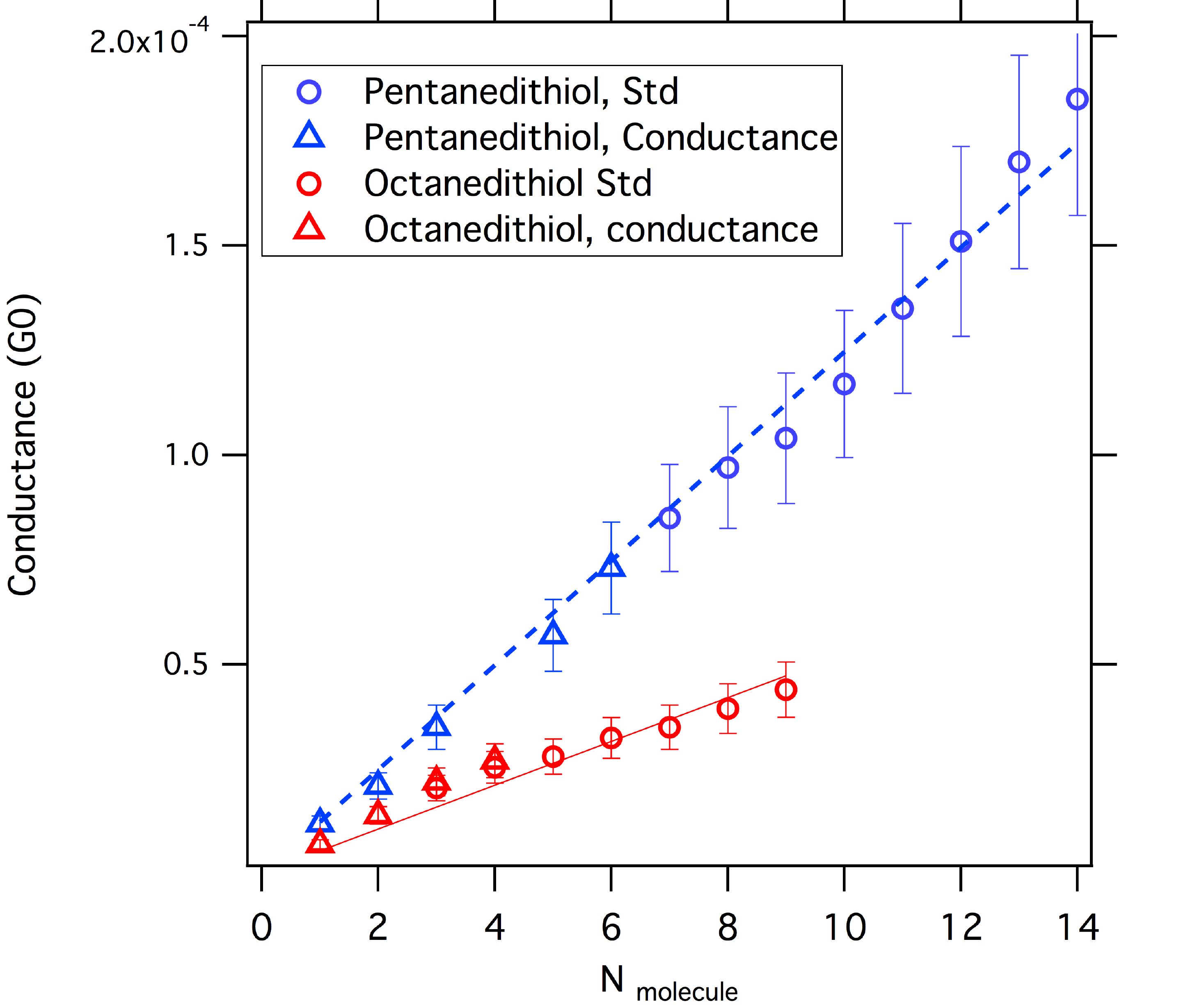}
\end{center}
\caption{ Conductance values corresponding to a discrete number of molecules in PDT and ODT contacts. The plot gathers data extracted from 
figure \ref{fig : conductance_values}(left) and \ref{fig : conductance_std}. It shows that the conductance values extracted from the two analyses are in excellent agreement. 
From linear fits we estimate the conductance of a single molecule to be $G_{PDT}= 1.25.10^{-5}\pm 0.19.10^{-5} G_0$ for PDT and 
$G_{ODT}= 5.27.10^{-6}\pm 0.22.10^{-6} G_0$ for ODT.}
\label{fig : conductance_values_final}
\end{figure}

The histograms in figure \ref{fig : conductance_histogram} exhibit peaks for up to 6 molecules in the contact. At higher conductances, 
fluctuations smear out the 
current jumps and peaks are difficult to assign in the histogram representation.
A further step is to analyze the conductance fluctuations. If we make the reasonable assumption that the conductance
fluctuations in the contact depend on the number of molecules, jumps in the fluctuations should be observed when the number of molecules 
varies. Moreover these jumps should be evenly spaced in conductance, like the peaks in the conductance histograms. This is the main feature of 
figure \ref{fig : conductance_std}, where the arrows indicate these conductance fluctuation jumps which, as expected, are evenly spaced in conductance. 

Figure \ref{fig : conductance_values_final} is a plot gathering conductance values corresponding to the discrete number of molecules for PDT and ODT, from conductance 
and standard deviation analysis. It shows an excellent agreement for conductance values extracted by both methods. 
From it, we more robustely estimate the conductance of a single molecule to be $G_{PDT}= 1.25.10^{-5}\pm 0.19.10^{-5} G_0$ for PDT, 
and $G_{ODT}= 5.27.10^{-6}\pm 0.22.10^{-6} G_0$ for ODT, from the spontaneous evolution of molecular contacts containing up to 9 molecules for 
ODT, and 14 for PDT. To our knowledge, the litterature offers no analysis constructed from raw data.

The conductance values, while in quantitative agreement with those previously obtained from spontaneous molecular connections \cite{Haiss2004,Haiss2006}, 
are lower than those generally extracted from STM-BJ experiments  on alkanedithiols (see e.g. \cite{Li2008, Gonzalez2008}).
This can be attributed to the superpositioning of a tunneling channel onto the molecular channel in STM-BJ experiments, making it tricky 
to compare absolute conductance values.
If instead we compare conductance ratios for PDT and ODT, we obtain a ratio $G_{PDT} / G_{ODT} = 2.37\pm 0.46$ in our experiments,
 in disagreement with the conductance ratio of $20\pm6.1$ reported in the the pioneering work of N.J. Tao and co-workers \cite{Xu2003a}.

Conductance measurements are finely modeled with a simple tunneling model for saturated molecules such as alkanedithiols, and 
the conductance is expressed as $G \simeq \exp(-\beta.N)$, with $\beta$ a decay constant expressed per carbon atom in the molecular backbone, and a prefactor taking into account the contact resistance 
(estimated to 6 k$\Omega$ for alkanedithiols anchored to gold electrodes \cite{Liu2008}). 
A value of $\beta = 1.0\pm0.1$  is often reported leading to the conductance ratio of $20\pm6.1$ in \cite{Xu2003a}.
 
We now discuss this apparent discrepancy, which we attribute to the fact that the conductance values reported here represent an 
 averaging over a conformer population at thermal equilibrium of the same molecule.

 Alkane molecules present different conformers from 
 \textit{all-trans} (conformer of lower energy) to \textit{all-gauche} con{-}formations (conformer of higher energy). At thermal 
 equilibrium, the fractional population of different conformers follows a Boltzmann's distribution. 
 For alkanes in general, the activation energy for conformer interconversion (corresponding to the rotation of a 
 methyl group around a carbon-carbon bond) is about 14kJ.mol$^{-1}$ (0.15 eV per molecule).  When measuring the conductance of a 
 molecule at room temperature and at a millisecond time scale following our protocol, the conductance is therefore a time average 
 of an ensemble of conformers.
 
 \begin{figure*}	
\begin{center}
\includegraphics[width=100mm]{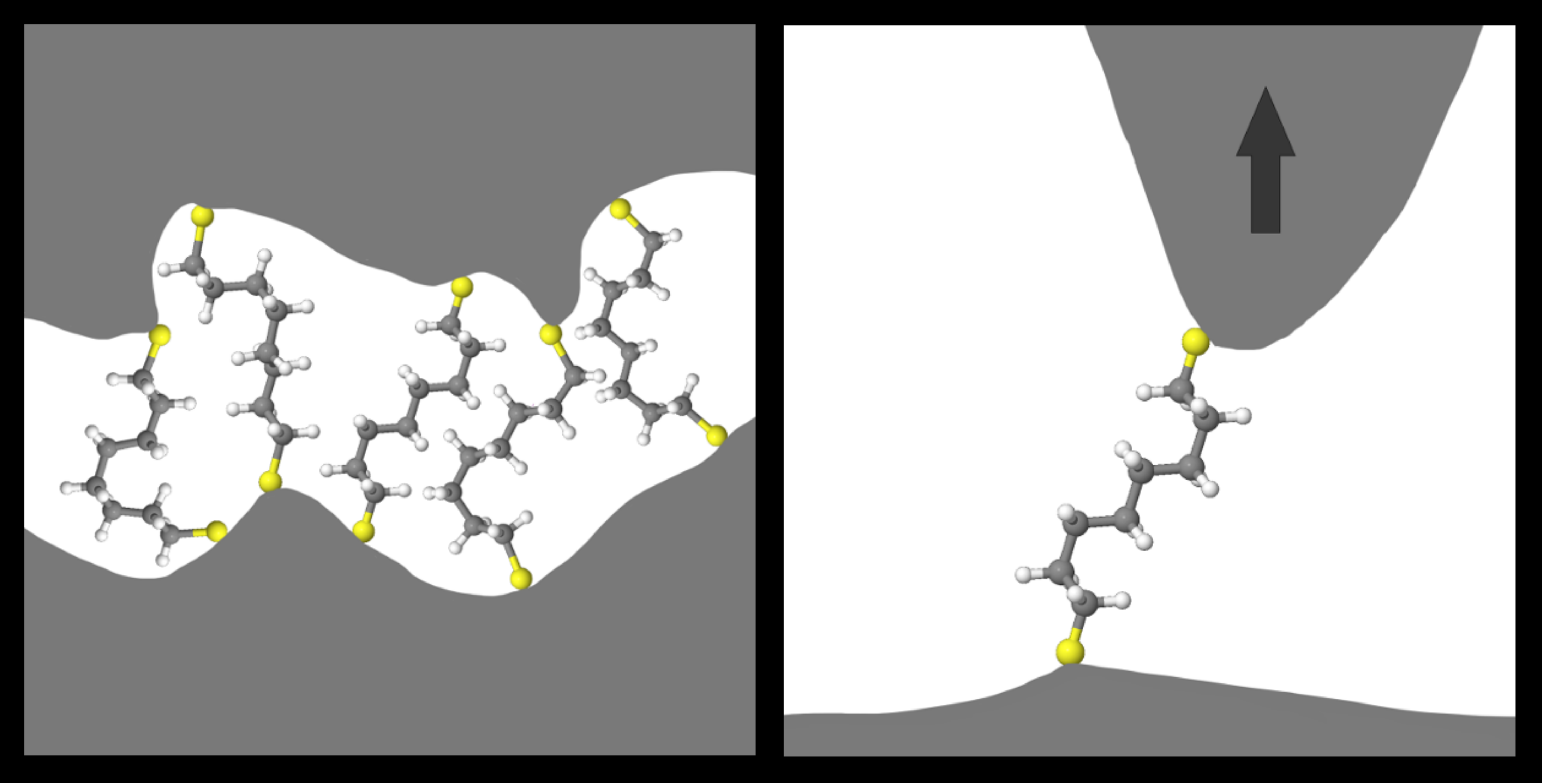}
\end{center}
\caption{Schematic illustration of conductance measurements in an MCBJ (left) and a STM-BJ (right) setup. While the traction of the anchored molecule 
by the STM tip favors the \textit{all-trans} conformers of alkanedithiol molecules in STM-BJ measurements, the morphology of the MCBJ electrodes 
allows the coexistence of multiple conformers of the molecules}
\label{fig : molecules}
\end{figure*}

  The PDT molecule presents 4 conformers, from the \textit{all-trans}, with the largest distance between 
 terminal groups, to the most folded conformer with the smallest. The length difference between the terminal groups for these two 
 conformers is $\Delta d=0.15$nm. If we make the simplest approximation that, for different alkanedithiols or different conformers of 
 the same alkanedithiol, 
 the barrier medium is the same,  and its width is related to the distance between terminal groups, the \textit{all-trans} conformer 
 has the largest barrier, while the \textit{all-gauche} has the smallest. If we calculate a tunneling current through a one-dimensional 
 rectangular barrier described by the decay constant $\kappa$ in nm$^{-1}$
 \begin{equation}
  \kappa = \left( \frac{2m^*_e \phi}{(h/2\pi)^2} \right)^{1/2} 
 \end{equation}

 where $h$ is the Planck's constant, $\phi$ the barrier height (in J) and $m^*_e$ the effective electron mass, taking $\phi=5$eV and 
 $m^*_e$ as the rest mass of the electron, we obtain for PDT a conductance ratio from the shortest to the longest conformer of 40.  
 Tunneling through the minimum energy configuration (i.e. all trans conformer) is unfavorable compared to 
 higher energy conformations potentially offering higher tunneling probabilities. Therefore, the contribution of each conformer 
 to the measured conductance is not directly proportional to its energy or probability of being occupied.
 
 This has been shown by W. Haiss and co-workers \cite{Haiss2006}, through the experimental and theoretical study of the temperature dependence of 
 single alkanedithiol molecule conductance. The dependence (conductance increases with temperature) is explained, using a simple 1D tunneling barrier, 
 by the change in the distribution between molecular conformers induced by temperature changes.
 From their results at room temperature, we can calculate a conductance ratio $G_{PDT} / G_{ODT} = 4\pm 1.3$ in excellent quantitative agreement with 
 our values.

In STM-BJ experiments, molecules are stretched during measurements, as illustrated in figure \ref{fig : molecules}. 
In these conditions the conformer populations are no longer in thermal equilibrium, and thus, \textit{all-trans} conformers are expected to be favored, 
leading to a higher $G_{PDT} / G_{ODT}$ ratio. Moreover, the contribution of a parallel tunneling channel in STM-BJ experiments may also increase 
the discrepancy with our results. 

\section{Conclusion and Perspectives}

We presented an original method to estimate the conductance of a single molecule anchored to metallic electrodes, by measuring the 
conductance of a molecular contact whose evolution is thermally driven. Spontaneous connections / deconnections of molecules within the 
contact are evidenced by abrupt conductance jumps at a millisecond time scale. The most probable configurations of the contact lead to
conductance values that are linearly spaced. We attribute these values to the opening / closing of conductance channels as a result of
discrete variation in the number of molecules in the contact. We estimated the conductance of a single molecule from a statistical analysis 
of raw conductance recordings on molecular contacts containing up to 14 molecules. While conceptually simple, this experiment 
relies on the outstanding mechanical stability of our MCBJ set-up, as the distance between the metallic electrodes must be kept 
constant during the measurements. 

We obtained conductance values for two molecules PDT and ODT,  using a robust statistical analysis method which 
removes the tunneling contribution of the electrodes from the net molecular conductance. In contrast 
to STM-BJ experiments, we measured the time-averaged conductance of an ensemble of conformers in thermal equilibrium at room temperature. 

Discrete variation in conductance and conductance standard deviation is observable for up to 9 molecules in the contact for ODT, and 14 for PDT. 
If we suppose that the molecules are packed in the contact with a typical distance between 
sulfur groups of 0.5nm \cite{Schreiber2000}, we can estimate the lateral extension of the larger aggregates to be in the range 1.5-3 nm.  
For larger aggregates, discrete variation in conductance is no longer visible, probably hidden by conductance 
fluctuations larger than the contribution of a single molecule. 
While quantitative noise analysis is far beyond the scope of this article, the measurement technique we detailed here opens the way to such 
studies on nanocontacts of tunable cross section, which are rarely reported \cite{Wu2008}. 
Its use promises new insights into the dynamic and transport mechanisms of nanometric-sized 
molecular contacts at room temperature.

\section{acknowledgments}
This work was partially funded by the ANR grant FOST, ANR-12-BS10-01801. We would like to thank Ms M. Sweetko for careful proof reading 
and English revision, Mr M. Lagaize for technical drawings and Ms T. Klein for artistic drawings.


\section*{References}

\bibliographystyle{unsrt}

\end{document}